\documentclass[preprint,12pt]{elsarticle}

\usepackage{amssymb}
\usepackage{amsmath}
\usepackage{bm}

\usepackage{siunitx}
\usepackage[version=4]{mhchem}
\usepackage{makecell}
\usepackage{multirow}

\usepackage[linesnumbered,lined,boxed,commentsnumbered]{algorithm2e}

\usepackage{enumitem}
\usepackage{subcaption}

\usepackage{hyperref}

\usepackage{nomencl}
\makenomenclature
\renewcommand\nomgroup[1]{%
  \item[\emph{
  \ifstrequal{#1}{A}{Acronyms}{%
  \ifstrequal{#1}{B}{Roman symbols}{%
  \ifstrequal{#1}{C}{Greek symbols}{%
  \ifstrequal{#1}{E}{Superscripts}{
  \ifstrequal{#1}{D}{Subscripts}{
  \ifstrequal{#1}{O}{Other symbols}{%
  }}}}}}%
}]}

\newcommand{\nomRoman}[1][]{\nomenclature[B,#1]}
\newcommand{\nomGreek}[1][]{\nomenclature[C,#1]}

\newcommand{\nomSub}[1][]{\nomenclature[D,#1]}
\newcommand{\nomAcro}[1][]{\nomenclature[A,#1]}

\journal{Applied Energy}

\begin{document}

\begin{frontmatter}



\title{Standby efficiency and thermocline degradation of a packed bed thermal energy storage: An experimental study}


\author[IET]{Paul Schwarzmayr\corref{corrauthor}}\ead{paul.schwarzmayr@tuwien.ac.at}
\author[IET]{Felix Birkelbach}
\author[IET]{Heimo Walter}
\author[IET]{René Hofmann}

\cortext[corrauthor]{Corresponding author}

\affiliation[IET]{organization={Institute for Energy Systems and Thermodynamics, TU Wien},
            addressline={Getreidemarkt 9}, 
            city={Vienna},
            postcode={1060}, 
            country={Austria}}



\begin{abstract}
The waste heat potential from industrial processes is huge and if it can be utilized it may contribute significantly to the mitigation of climate change. A packed bed thermal energy storage system is a low-cost storage technology that can be employed to enable the utilization of waste heat from industrial processes. This system can be used to store excess heat and release this energy when it is needed at a later time. To ensure the efficient operation of a packed bed thermal energy storage its characteristics in standby mode need to be studied in great detail. In the present study, the standby efficiency and thermocline degradation of a lab-scale packed bed thermal energy storage in standby mode is experimentally investigated for different flow directions of the heat transfer fluid during the preceding charging period. Results show that, for long standby periods, the standby efficiency is significantly affected by the flow direction. The maximum entropy generation rate for a 22-hour standby process with the flow direction of the heat transfer fluid from bottom to top in the preceding charging process is twice as high as for the same process with the reverse flow direction. Energy and exergy efficiencies are lower for the process with reverse flow direction by $5\%$ and $18\%$ respectively.\\
\end{abstract}

\begin{graphicalabstract}
\includegraphics[width=14cm]{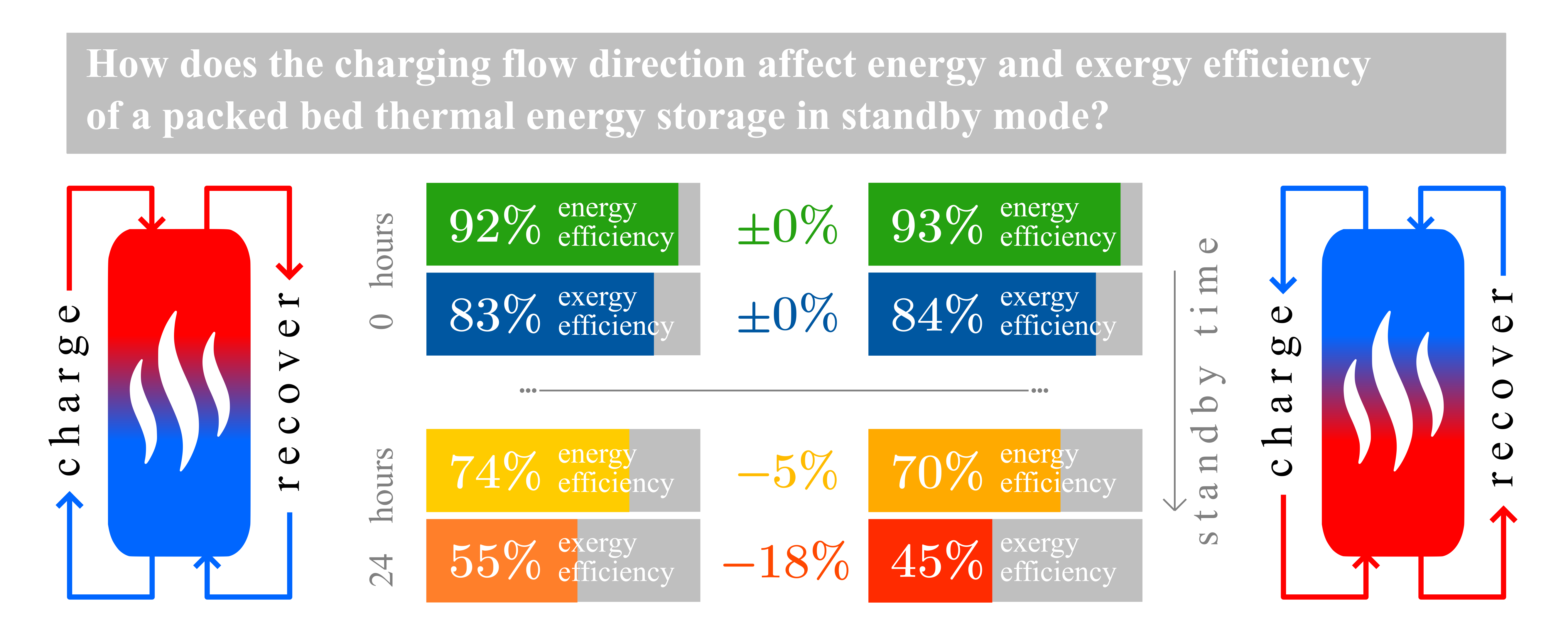}
\end{graphicalabstract}

\begin{highlights}
\item Standby mode of a packed bed thermal energy storage is experimentally investigated.
\item Energy and exergy efficiencies of max. $92\%$ and $83\%$ are measured.
\item Measurement data is processed by second law analysis.
\item Standby efficiencies depend on the air flow direction in the charging period.
\item The impact of natural convection on long standby periods is significant.
\end{highlights}

\begin{keyword}
packed bed \sep thermal energy storage \sep standby mode \sep stratification \sep thermocline degradation \sep energy/exergy efficiency
\end{keyword}

\end{frontmatter}



\section{Introduction}
\label{sec:sample1}
The recovery of waste heat in the energy intensive industry has great potential to make a significant contribution to the mitigation of climate change. More than two third ($\approx 70\%$) of the industry sectors' energy demand is for heat \cite{jouhara_editorial_2018}. For some sectors up to one half of this heat consumption is discharged as waste heat as stated by Papapetrou et al. \cite{papapetrou_industrial_2018}, who estimated the available waste heat potential for the EU per industrial sector, temperature level and country. A similar study was done by Panayiotou et al. \cite{panayiotou_preliminary_2017} who made an assessment off the waste heat potential in major European industries. In their study on the estimation of the globally available waste heat, Forman et al. \cite{forman_estimating_2016} concluded that 72\% of the global primary energy consumption is dissipated after conversion. Especially in the iron and steel industry heat is often wasted because of temporal discrepancies of heat demand and supply. Thermal energy storage (TES) technologies can be used to improve the efficiency of these discontinuous batch processes by closing the gap between heat demand and supply. Manente et al. \cite{manente_structured_2022} studied the integration of TES systems into industrial energy processes and found that the integration of a TES, either of sensible or latent type, results in an enhanced energy efficiency and an increased process output.\\

A simple but highly efficient type of TES is a packed bed thermal energy storage (PBTES). Comprehensive reviews on packed bed thermal energy storage systems were done by Gautam et al. (\cite{gautam_review_2020}, \cite{gautam_review_2020-1}) and Baoshan et al. \cite{xie_thermocline_2022}. PBTES systems are sensible TES systems which use a packed bed of rocks or other suitable materials as storage material. To charge a PBTES hot heat transfer fluid (HTF), in most cases air, passes through the bed delivering heat to the storage material. For the recovery of stored heat the flow direction inside the storage is reversed and cold HTF is used to extract heat from the storage material. Since the HTF makes direct contact with the packed bed a vast surface area is provided for the heat transfer inside the storage. This leads to very high power rates, a high temperature stratification inside the packed bed and therefore an overall enhancement of the system's efficiency. Temperature stratification means that the storage volume is separated in a hot and a cold region. These two regions are separated by a thin volume-slice, the thermocline, where the whole temperature gradient is concentrated. During charging, the hot HTF pushes the thermocline towards one end of the storage, expanding the hot region. During discharging the cold HTF pushes the thermocline in the reverse direction. With an ideal thermocline the discharging temperature of a PBTES theoretically is as high as the charging temperature for the whole process. In reality, irreversible effects like heat losses to the surrounding, heat conduction and natural convection inside the packed bed lead to a degradation of the thermocline and therefore to a reduced efficiency of the storage. An extensive review on experience feedback and numerical modeling of PBTES systems can be found in the work of Esence et al. \cite{esence_review_2017}, in which they pointed out the importance of maintaining high temperature stratification in a PBTES in order to maximize exergy efficiency and storage capacity. Similar conclusions were drawn by Stekli et al. \cite{stekli_technical_2013} who pointed out that the biggest techno-economic challenge of thermocline TES systems is thermocline degradation. An experimental study on multiple consecutive charging/discharging cycles was carried out by Bruch et al. (\cite{bruch_experimental_2014}, \cite{bruch_experimental_2017}) on a PBTES with rocks as storage material and thermal oil as HTF. Bruch et al. concluded, that for a well designed and well operated PBTES in dynamic operation, i.e. consecutive charging/discharging cycles, thermocline degradation is negligible. However, Okello et al. \cite{okello_experimental_2014} found that if the heat stored in a stratified storage is to be used after some time (12-24h) a significant degradation of the thermocline and therefore significant exergy losses can be observed as the system tries to reach thermal equilibrium.\\

Studies associated with the static operation, i.e. the standby mode, of TES systems such as the one from Okello et al. are rare. Mertens et al. \cite{mertens_dynamic_2014} numerically investigated a PBTES for concentrated solar power (CSP) applications, also considering the standby mode. Xu et al. \cite{xu_parametric_2012} carried out numerical studies on the standby mode of a PBTES with molten salt as HTF.
The most recent publications considering the static operation of a PBTES are from Rodrigues et al. \cite{rodrigues_stratification_2021} and Yang et al. \cite{yang_study_2019}. Rodrigues et al. numerically investigated the destratification after a charging process also considering laminar natural convection including the local thermal non-equilibrium assumption. The study from Yang et al. includes numerical investigation with experimental validation on both the dynamic and static operation of a pilot-scale PBTES test rig. Both studies only investigated the standby process after charging a TES from top, which will be the preferred option of course. Given that a hot region and a cold region coexist in the storage, buoyancy effects and therefore natural convection may occur in the tank. Charging from top will always be the first choice \cite{esence_review_2017} in order to minimize the destratification, especially during considerably long standby periods.\\

Future industrial energy systems will be highly flexible systems with fluctuating and hard-to-predict demand and supply trajectories. If TES systems are to be a part of future energy systems they need to be highly flexible and efficient as well. They need to deal with various demand and supply fluctuations and therefore should have the ability for different operation modes including static operation, where the TES is in standby for a longer period of time. The operational integration of a TES into industrial processes is often optimized by defining a unit commitment problem and solving it with methods like mixed integer linear programming. As discussed by Koller et al. \cite{Koller_milp_2019} the quality of the optimization results depends on the accuracy of the underlying models. In order to optimize the standby mode of a PBTES accurate models and hence a detailed understanding of its thermal behaviour is indispensable. In addition to these operational requirements there are further design requirements which modern TES systems have to meet. In their review on system and material requirements of TES systems Gasia et al. \cite{gasia_review_2017} identified, that besides thermo-physical properties of storage materials and economic requirements, technological requirements, which include the integration of TES systems into the facility, are of high importance. Cabeza et al. \cite{cabeza_1_2021} even stated, that one of the most important design criteria of TES systems is their integration into the whole application system. TES systems need to be integrated into existing energy systems with as little change to the existing system as possible. In some cases it can be necessary to charge the TES from the bottom and discharge it from the top because of construction limitations or economic considerations (e.g. modification costs). In this configuration the destratification during a standby mode may be different.\\

To the best of the authors' knowledge, no studies on the impact of the HTF flow direction during charging on the destratification and the standby efficiency of PBTES systems has been published. Therefore the aim of this work is to experimentally investigate the standby mode of a PBTES test rig for different HTF flow directions during the preceding charging process. The goal is to provide information to make informed decisions. So that industrial companies can decide whether and how TES systems can be integrated into existing processes, a basis for economic assessments and payback period calculations is needed. Specifically for PBTES, the storage capacity, the thermal power rate and the energy and exergy round-trip efficiency are key performance indicators and can be used as a basis for these calculations. The present study provides reference values for the mentioned key performance indicators during the standby mode for different operational conditions.\\

Section \ref{sec:methods} of this paper includes a presentation of the experimental setup used for the investigation as well as a description of the utilized measurement equipment, the sensor layout and the measurement control. In Section \ref{sec:analysis}, the first and second law are used for data analysis so that the internal effects in a PBTES during a standby period can be quantified. In the end of Section \ref{sec:analysis} the impact of measurement errors on the calculated quantities is presented and discussed shortly. The experimental data and the derived results are discussed in Section \ref{sec:results}. An interpretation of the entropy generation equation as well as reference values for energy and exergy efficiencies under different operational conditions are given and discussed extensively. Finally, Section \ref{sec:conclusion} provides a summary of the main results as well as a conclusion considering the consequences and the potential of the novel insights collected in this work.

\section{Material and Methods}
\label{sec:methods}

\begin{figure}
    \centering
    \includegraphics[width=9cm]{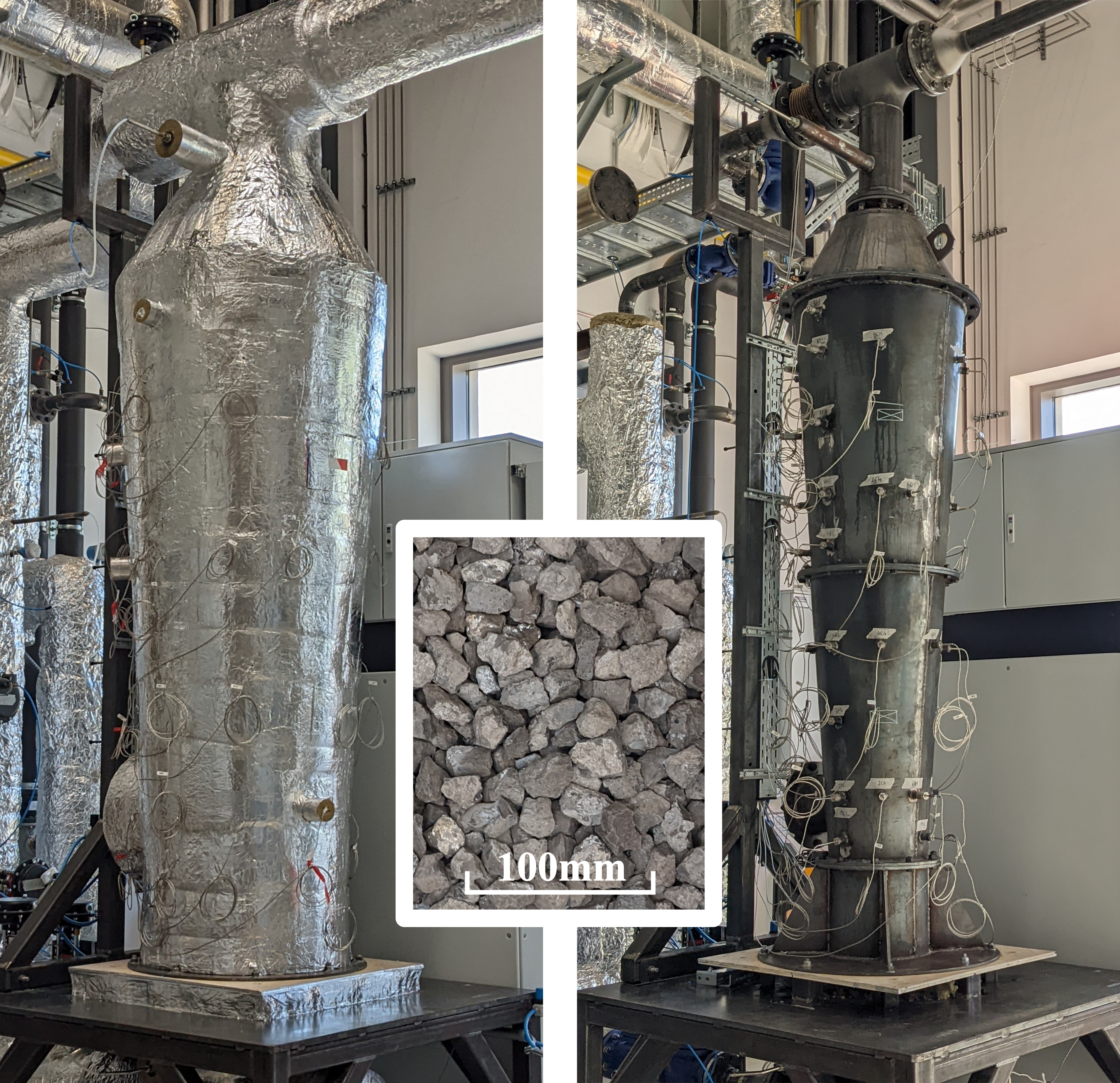}
    \caption{PBTES test rig at the laboratory of TU Wien: with thermal insulation (left), storage material (center), without thermal insulation (right)}
    \label{fig:testrig}
\end{figure}

The experimental setup for the present investigations consists of a lab-scale PBTES test rig filled with LD-slag as storage material. The LD-slag, a by-product from a steel producing process called the Linz-Donawitz(LD)-process, consists of irregular shaped, porous rocks (see Figure \ref{fig:testrig}). The geometric shape of the LD-slag leads to an enhanced heat transfer between the HTF and the storage material as well as homogeneous and even perfusion of the packed bed. The storage vessel is a vertically standing, conical steel vessel filled with the storage material. To minimize heat losses to the surrounding the TES is insulated with multiple layers of ceramic wool, rock wool and aluminum sheeting. Figure \ref{fig:testrig} shows the test rig before and after the thermal insulation is applied. The test rig is connected to an air supply unit (ASU) that provides air at temperatures between 20 and 400 °C and a mass flow between 100 and 400 kg/h.\\

\begin{figure}
    \centering
    \includegraphics[width=9cm]{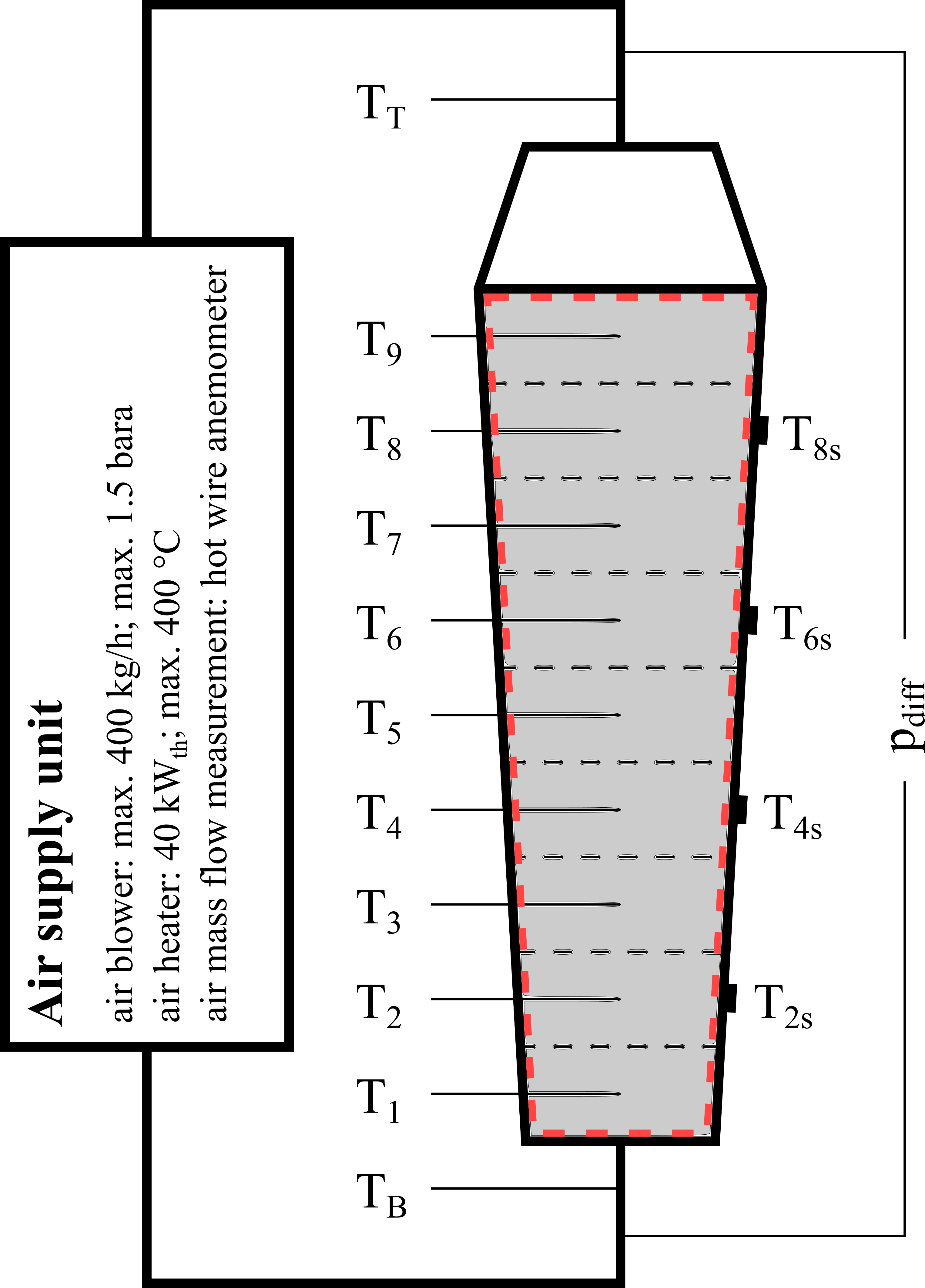}
    \caption{Storage instrumentation and discretization}
    \label{fig:pnid}
\end{figure}

Additionally, the experimental setup is equipped with several temperature and differential pressure sensors. In Figure \ref{fig:pnid} a simplified P\&ID-diagram including the equipment relevant for the present investigations is depicted. For detailed information about the sensor layout and the process control system the authors refer to their previous work \cite{schwarzmayr_development_2022}. Four-wire resistance temperature detectors (RTD) class AA (or 1/3 DIN) PT100 sensors were utilized for the temperature measurements. The test rig is equipped with two PT100 temperature sensors ($T_{\mathrm{T}}$ and $T_{\mathrm{B}}$), that measure the HTF temperature at the in- and outlet of the storage vessel, as well as nine PT100 temperature sensors ($T_{\mathrm{1}}, T_{\mathrm{2}}, ..., T_{\mathrm{9}}$) inside the packed bed and eight PT100 temperature sensors($2\times T_{\mathrm{s2}}, 2\times T_{\mathrm{s4}}, 2\times T_{\mathrm{s6}}, 2\times T_{\mathrm{s8}}$) on the lateral surface of the storage between the steel vessel and the first layer of insulation. For data analysis the storage volume is discretized into nine vertical sections as it is shown in Figure \ref{fig:pnid}. Each of these sections is equipped with a temperature sensor located at the center axis of the storage vessel. The eight sensors on the lateral surface of the TES are mounted in pairs opposite of each other in every second section of the TES. In Figure \ref{fig:pnid} only one sensor of each pair is shown. Measurements for the HTF massflow, humidity and the temperature of the surrounding are provided by the ASU. For a measured value of $\SI{300}{\celsius}$, the measurement error of the temperature sensors including all steps of signal conversion do not exceed $\pm\SI{0.6}{\celsius}$. The massflow of the heat transfer fluid is measured with a hot-wire anemometer with a maximum measurement error of $4\%$ with respect to the measured value.  Details on the test rig geometry, properties of the storage material and operational parameters used for the experiments are given in Table \ref{tab:param}.\\
\begin{table}
\begin{center}
\caption{Summary of parameters: Test rig geometry, data/properties of storage material, operational parameters}
\label{tab:param}
\begin{tabular}{l l} 
\hline
\multicolumn{2}{l}{\textbf{Test rig}}\\ \hline
Storage type                    &   \makecell[l]{vertical packed bed\\ thermal energy storage} \\
Tank material         & P295\\
Tank diameter         & \makecell[l]{\SI{200}{\milli\meter} (bottom)\\ \SI{500}{\milli\meter} (middle)\\ \SI{700}{\milli\meter} (top)}\\
Tank height         & \SI{2050}{\milli\meter}\\
Storage volume         & \SI{0.405}{\cubic\meter}\\
Thermal insulation                    & \makecell[l]{\SI{100}{\milli\meter} of ceramic wool\\ ($\lambda = \SI{0.055}{\watt\per\meter\per\kelvin}$)\\ \SI{80}{\milli\meter} of rock wool\\ ($\lambda = \SI{0.04}{\watt\per\meter\per\kelvin}$)\\ \makecell[l]{\SI{0.1}{\milli\meter} of aluminum\\ sheeting}}\\
\hline
\multicolumn{2}{l}{\textbf{Storage material}}\\ 
\hline
\makecell[l]{Type of storage\\ material}           & \makecell[l]{LD-slag (irregular\\ shaped, partly porous)} \\
\makecell[l]{Composition of\\ storage material}         & \makecell[l]{\ce{CaO} (24 - 49 \%),\\ \ce{SiO2} (6 - 37 \%),\\ \ce{Fe_{x}O_{y}} (10 - 36 \%),\\ \ce{MgO} (0 - 13 \%),\\ \ce{Al2O3} (0 - 7 \%),\\ \ce{Cr2O3} (0 - 0.55 \%)}\\
Grain size                               & \SIrange{16}{32}{\milli\meter}\\
Particle density          & \SI{3800}{\kilo\gram\per\cubic\meter}\\
Porosity                & 0.42\\
Bulk density          & \SI{2200}{\kilo\gram\per\cubic\meter}\\
Specific heat capacity       & \SI{900}{\joule\per\kilo\gram\per\kelvin}\\ \hline 
\multicolumn{2}{l}{\textbf{Operating parameters}}\\ \hline
\makecell[l]{Mass of storage\\ material}          & \SI{891}{\kilo\gram}\\
Air mass flow                     & \SI{200}{\kilo\gram\per\hour}\\
Charging temperature              & \SI{300}{\celsius}\\
Recovery temperature              & \SI{20}{\celsius}  \\
\makecell[l]{Mean superficial\\ air velocity}     & \SI{1}{\meter\per\second}  \\\hline
\end{tabular}
\end{center}
\end{table}

For the investigations carried out in this work experiments for different standby times are conducted. The experiments start with an empty storage (temperatures inside the storage between $\SIrange{20}{25}{\celsius}$) before it is charged for two hours. After the charging process is finished, the TES is in standby for different time periods (\SI{0.5}{\hour} and \SI{22}{\hour}) before the stored heat is recovered again. The cut-off temperature for recovery is set to $\SI{70}{\celsius}$. This procedure is repeated for both charging the TES from the top, which will be referred to as the FWD-case, and from bottom, which will be referred to as the REV-case.

\section{Theory and Calculations}
\label{sec:analysis}
To make statements about the impact of the charging direction of a PBTES on thermocline degradation and the standby efficiency temperature data from the experiments is used. The data analysis is done in a way, that effects from heat losses to the surrounding and due to heat transfer inside the packed bed can be examined separately. An elegant way to achieve this is to make use of the entropy generation equation (Equation \eqref{eq:s_bal}), i.e. the expression of the second law of thermodynamics for closed systems. The left side of this equation represents the change in system entropy. The right side is the sum of the entropy introduced into the system due to heat crossing the system boundary and the entropy generated within the system.\\

If this equation is applied to a system with suitable system boundaries, the entropy generation term is influenced by irreversible effects due to heat transfer inside the packed bed only. For the present work, a system with boundaries that enclose nothing more than the packed bed itself is chosen (marked by the red dashed line in Figure \ref{fig:pnid}). This way only irreversible effects from heat transfer inside the packed bed contribute to the entropy generation term and irreversibilities due to heat transfer to the surrounding are not included. The entropy generation equation for the chosen system can be written as\\
\begin{equation}
    \sum_{i=1}^{n} \mathrm{d}S_{\mathrm{sys}, i}(t) = \frac{\delta Q_{\mathrm{t/b}}(t)}{T_{\mathrm{t/b}}(t)} + \sum_{i=1}^{n}\frac{\delta Q_{\mathrm{lat}, i}(t)}{T_{i}(t)} + \delta S_{\mathrm{gen}}.
    \label{eq:s_bal}
\end{equation}

The change in system entropy (left hand side of Equation \eqref{eq:s_bal}) can be calculated from the measured temperature data using the equation of state in Equation \eqref{eq:ds}.\\
\begin{equation}
    \mathrm{d}S_{\mathrm{sys}, i}(t) = m_{i}\,\int\limits_{T_{i}(t)}^{T_{i}(t+dt)}\frac{c(T)}{T}dT
    \label{eq:ds}
\end{equation}

For the next term in Equation \eqref{eq:s_bal}, the entropy introduced into the system due to heat crossing the system boundary, the heat losses to the surrounding need to be calculated. With a suitable correlation for the Nusselt number for natural convection on a vertically standing cylinder \cite{merker_konvektive_1987}, the lateral heat loss rate to the surrounding for each of the nine vertical sections of the TES can be calculated using Equation \eqref{eq:heattransfer}.\\
\begin{equation}
    \dot{Q}_{\mathrm{lat}, i}(t) = k\,A_{\mathrm{lat}, i}\,\left(T_{\mathrm{amb}}-T_{\mathrm{s}, i}\right)
    \label{eq:heattransfer}
\end{equation}

The heat transfer coefficient $k$ in Equation \eqref{eq:heattransfer} is calculated from geometry data, the heat transfer coefficient to the surrounding and thermo-physical properties of the insulation material (see Table \ref{tab:param}). The thermal conductivity of the insulation material is assumed constant for the whole temperature range. For $T_{\mathrm{s}, i}$ the measurements of the temperature sensors on the interface between the steel vessel and the first layer of insulation are used.\\

Considering the standby process of the TES, where the only heat flows crossing the system boundaries are heat losses to the surrounding, the energy balance for the packed bed can be written as\\
\begin{equation}
    \sum_{i=1}^{n} \mathrm{d}U_{\mathrm{sys}, i}(t) = \delta Q_{\mathrm{t/b}}(t) + \sum_{i=1}^{n} \delta Q_{\mathrm{lat}, i}(t).
    \label{eq:yhat}
\end{equation}

The change in the systems internal energy equals the sum of all heat losses. This equality allows to calculate the heat losses on the top and bottom surfaces of the packed bed. Finally, Equation \eqref{eq:s_bal} can be rearranged to calculate the entropy generation inside the system.\\

Another way to make statements about the impact of the charging direction on thermocline degradation and standby efficiency is to calculate energy and exergy round-trip efficiencies for the FWD and REV case and compare them to each other. Therefore characteristic parameters like the thermal power rate of the storage and the stored/recovered thermal energy for the charging and discharging processes need to be calculated. The power rate is determined from the energy balance for the HTF flow through the packed bed:\\
\begin{equation}
    \dot{Q}(t) = \dot{m}(t)\,\left[c_{p}(T_{\mathrm{out}}(t))\,T_{\mathrm{out}}(t)-c_{p}(T_{\mathrm{in}}(t))\,T_{\mathrm{in}}(t)\right]
    \label{eq:qdot}
\end{equation}

The stored/recovered energy is defined by the integral of the power rate over a certain time interval:\\
\begin{equation}
    U(t) = \int\limits_{t_{\mathrm{0}}}^{t}\dot{Q}(t)\,\mathrm{d}t
    \label{eq:utes}
\end{equation}

The rate of exergy in-/output to and from the system is calculated using another expression of the second law of thermodynamics as it is written in Equation \eqref{eq:bdot} applied to the HTF flow through the packed bed. The stored/recovered exergy $B(t)$ can be determined analogously to Equation \eqref{eq:utes} by substituting $\dot{Q}(t)$ with $\dot{B}(t)$, where\\
\begin{equation}
    \dot{B}(t) = \dot{Q}(t) - T_{\mathrm{ref}}\,\dot{m}(t)\,\int\limits_{T_{\mathrm{in}}}^{T_{\mathrm{out}}}\frac{\,c_{p}(T)}{T}\,\mathrm{d}T.
    \label{eq:bdot}
\end{equation}

Finally, the energy efficiency $\eta_{\mathrm{e}}$ and the exergy efficiency $\eta_{\mathrm{b}}$ can be calculated as the ratios of the recovered quantities to the stored quantities:\\
\begin{equation}
    \eta_{\mathrm{e}} = \frac{U_{\mathrm{recover}}}{U_{\mathrm{charge}}}\, , \,\eta_{\mathrm{b}} = \frac{B_{\mathrm{recover}}}{B_{\mathrm{charge}}}
    \label{eq:eta}
\end{equation}

The gaussian law of error propagation is used to estimate the impact of measurement errors on the derived quantities. For calculations where only measurements from temperature sensors are used, the resulting uncertainties are insignificant compared to the measured values. Calculations were the HTF mass flow is included show slightly higher errors in the calculated quantities. In Equation \eqref{eq:error} the law of error propagation applied to the energy balance in Equation \eqref{eq:qdot} is given. For all experiments the resulting errors in the calculated heat flow rate $\dot{Q}$ are well below $4.5\%$ with respect to the measured value. These uncertainties do not have an impact on the quality of the results in the present work.
\begin{equation}
                \delta \dot{Q}^2=\left(\frac{\partial \dot{Q}}{\partial \dot{m}}\right)^2\,\delta\dot{m}^2+\left(\frac{\partial \dot{Q}}{\partial h_\mathrm{{in}}}\right)^2\,\delta h_\mathrm{{in}}^2+\left(\frac{\partial \dot{Q}}{\partial h_\mathrm{{out}}}\right)^2\,\delta h_\mathrm{{out}}^2
                \label{eq:error}
\end{equation}
\begin{equation}
                \delta h_i=c_p\left(T_i\right)\,\delta T_i
                \label{eq:errorad}
\end{equation}

\section{Results and Discussion}
\label{sec:results}
\begin{figure}
    \centering
    \includegraphics[width=9cm]{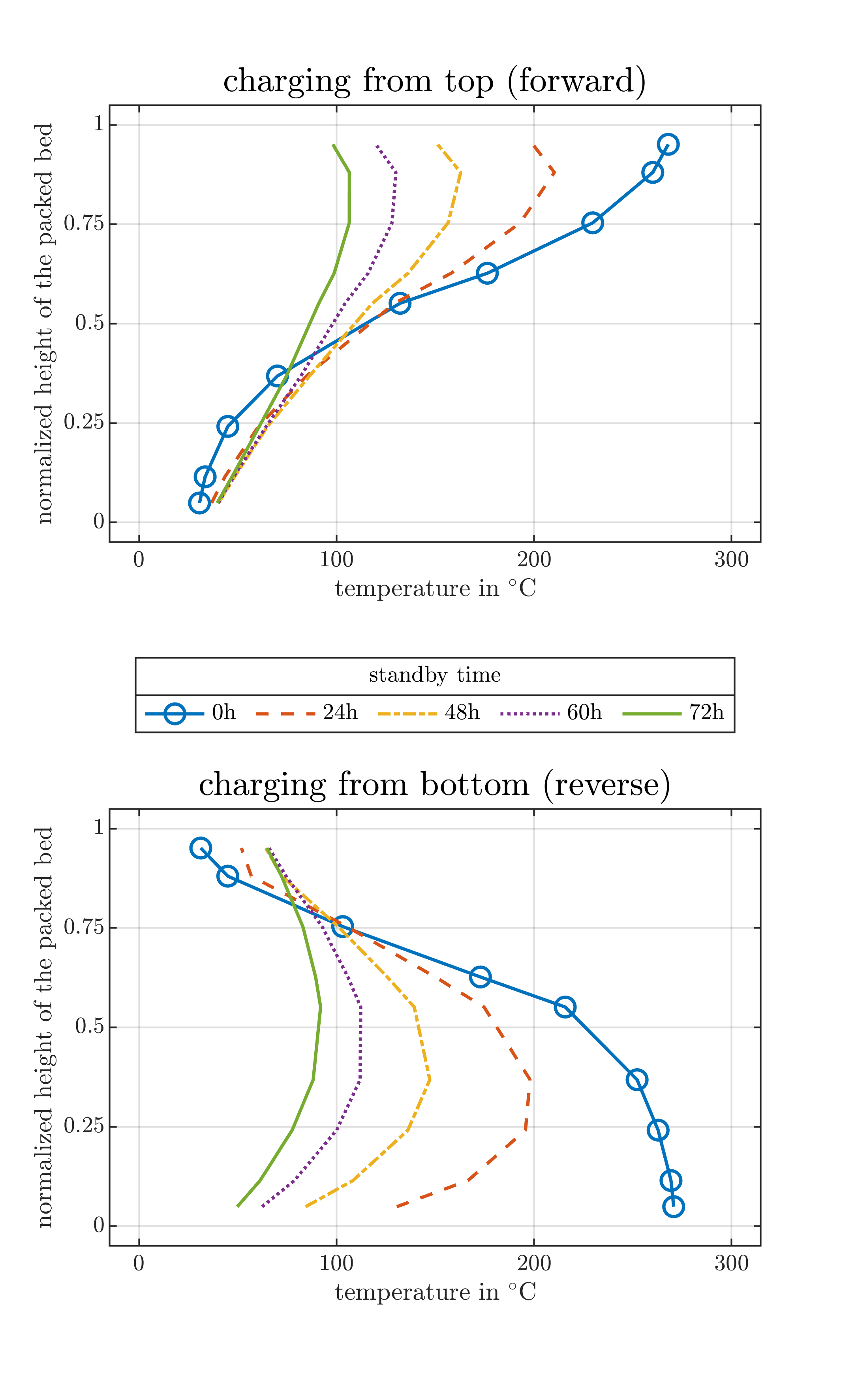}
    \caption{Thermocline degradation for charging from top (top plot) and charging from bottom (bottom plot) during a standby period}
    \label{fig:thermocline}
\end{figure}

The data collected in the experiments described in the end of Section \ref{sec:methods} are used to make both qualitative and quantitative statements on the standby efficiency and thermocline degradation of the examined PBTES test rig. Figure \ref{fig:thermocline} shows the degradation of the thermocline inside the test rig after charging for two hours in FWD direction (top graph) and in REV direction (bottom graph). The thermocline directly after the charging process (standby time = 0h) is represented by the solid blue line. The $\circ$-markers show the vertical position of the temperature sensors. For both flow directions the thermoclines have a similar shape after charging. The slight difference in the position of the thermoclines is due to the conical shape of the storage vessel and therefore a slightly uneven mass distribution of the storage material.\\

After some time in standby the temperatures in the storage decrease and the thermoclines start to degrade as it is shown in Figure \ref{fig:thermocline}. For the "charging from top"-experiment, i.e. the FWD-case, the thermocline flattens and preserves its shape for the whole standby period. The slight bend in the thermocline at the hot (top) end of the packed bed is due to heat losses on the top surface. For the "charging from bottom"-experiment, i.e. the REV-case, the thermocline flattens too but, in addition, significantly changes its shape. Again, at its hot end the thermocline develops a bend which is due to heat losses on the bottom surface of the packed bed. But, this time this bend is much more pronounced. That is because in the first case the thermocline is self-stabilizing due to buoyoncy effects and in the second case it is not. In the FWD-case the heat losses at the top surface are compensated with heat transported to the top by natural convection and in the REV-case no effects that could compensate the heat losses on the bottom surface are present. In addition to this bend a faster degradation of the thermocline in the REV case can be noticed. In contrast to the FWD case the temperature at the cold end of the packed bed rises and the maximum temperature inside the storage moves towards the cold end. These observations suggest that natural convection in the packed bed may have a noticeable impact on the standby mode of a PBTES.\\

\begin{figure}
    \centering
    \includegraphics[width=9cm]{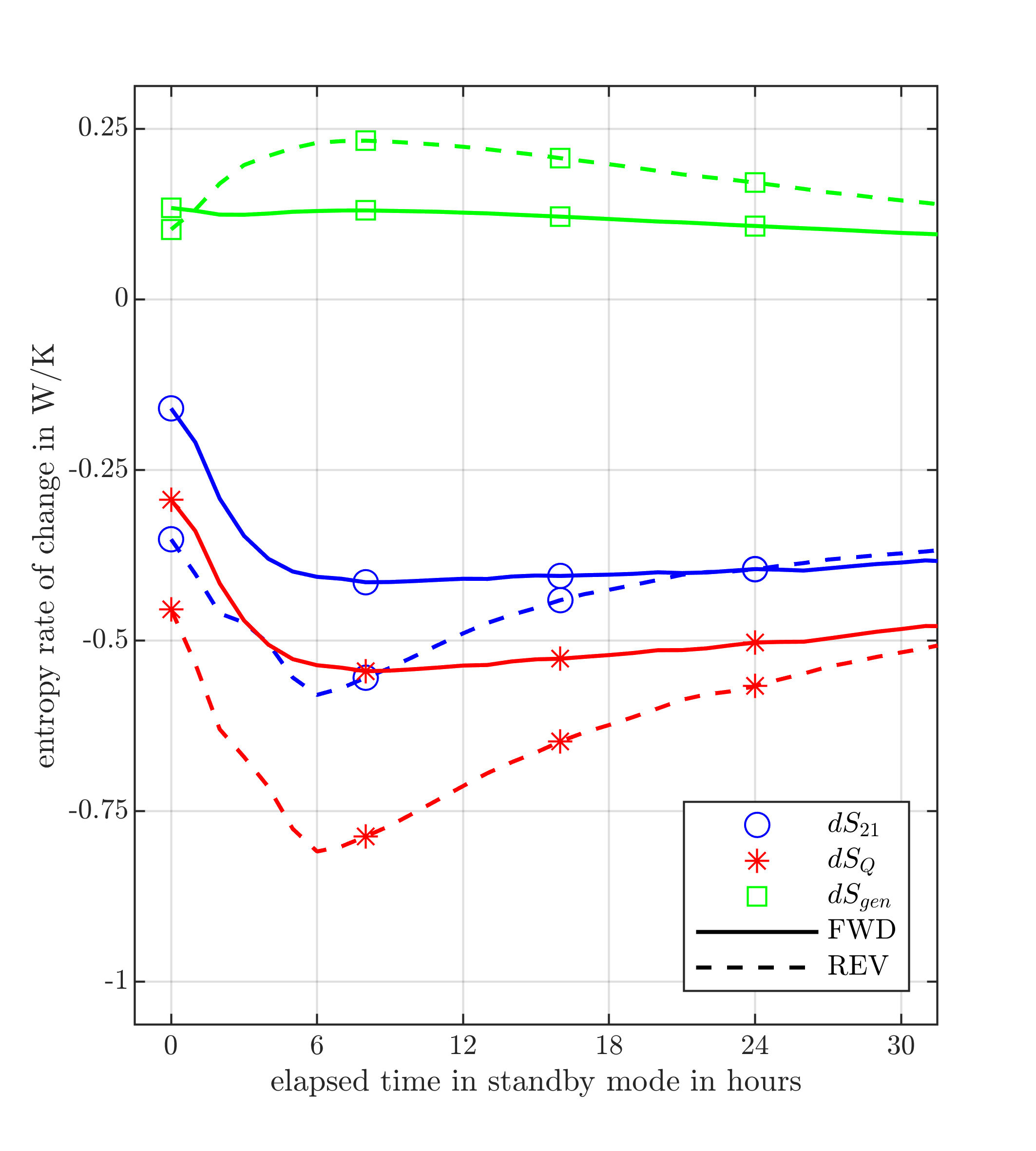}
    \caption{Entropy generation for a standby period of $\approx \SI{33}{\hour}$}
    \label{fig:entropy}
\end{figure}

For the quantification of these effects the entropy generation equation in Equation \eqref{eq:s_bal} is used. The result of these evaluations are visualized in Figure \ref{fig:entropy}. This graph includes the three terms from Equation \eqref{eq:s_bal} plotted as separate curves for a standby period of the test rig after 2h of charging for both the FWD and REV case. Blue lines with $\circ$-markers show the rate of change in system entropy $\dot{S}_{sys}$, red lines with $\ast$-markers show the entropy flow rates due to the heat flow rate across the system boundary $\dot{S}_{Q}$ and the green lines with $\square$-markers show the entropy generation rate inside the system $\dot{S}_{gen}$. The x-axis represents the elapsed time in standby mode. Notice that the absolute values of $\dot{S}_{sys}$ and $\dot{S}_{Q}$ are higher for the REV-case. This is because of the higher heat losses to the surrounding due to a higher surface-to-volume ratio in the bottom half of the storage. Nevertheless, these additional heat losses do not have an impact on the results and conclusions drawn in the present work as it will be discussed later in this section.\\

Much more interesting is, that there is also a significant difference in the entropy generation rate term $\dot{S}_{gen}$. As explained in Section \ref{sec:analysis} this term is influenced only by irreversible effects inside the packed bed, which could be heat transfer due to conduction, heat transfer due to natural convection and heat transfer due to radiation. Heat transfer due to radiation is negligible because of the vanishingly small temperature differences inside the packed bed. In the FWD case the hot zone in the storage is above the cold zone which makes the effects of natural convection insignificant too. Therefore the entropy generated in the FWD-case is only caused by irreversible effects due to heat conduction in the solid storage material. Because the heat conduction is not affected by buoyancy forces the same amount of entropy is generated in the REV-case due to heat conduction as well. Considering Figure \ref{fig:entropy} this is true for the first few minutes of the standby period. However, in the REV-case, the entropy generation term increases until it reaches a maximum with two times the value of the FWD-case after about seven hours in standby. This extra entropy production can only be explained by irreversible effects caused by heat transfer due to natural convection inside the packed bed. In addition it can be determined, that it takes the system about seven hours to establish a fully developed air circulation inside the packed bed.\\

These results are also supported by evaluating the efficiencies defined in Equation \eqref{eq:eta}. If one considers the ratio $\eta_{\mathrm{b}}/\eta_{\mathrm{e}}$, where $\eta_{\mathrm{b}}$ is the exergy round-trip efficiency and $\eta_{\mathrm{e}}$ the energy round-trip efficiency, the value of this ratio would equal one for both the FWD- and the REV-case, if no irreversible effects like heat transfer were present inside the system even if heat losses to the surrounding were different. It would be less than zero but still the same value for both cases, if irreversible effects were present to the same extent, again, even if heat losses to the surrounding are different. However, if the value of this ratio is different for the FWD-case and the REV-case, more irreversible effects happen in the case with the smaller ratio $\eta_{\mathrm{b}}/\eta_{\mathrm{e}}$. And that is exactly what is found by the evaluation of the measured data (see Table \ref{tab:ratio}). The argumentation of why these additional irreversibilities can only be caused by heat transfer due to natural convection is the same as before.\\

\begin{table}
\begin{center}
\caption{energy- and exergy efficiencies for experiments with different charging directions (FWD and REV) and standby times (\SI{0.5}{\hour} and \SI{22}{\hour})}
\label{tab:ratio}
\begin{tabular}{c | c | c c | c} 
\hline
 &  & FWD & REV & $\Delta\eta_{\mathrm{rel}}$ \\ \hline
\multirow{3}{*}{\makecell{\SI{0.5}{\hour}}} & $\eta_{\mathrm{e}}$ & 0.92 & 0.93 & $\approx0\%$ \\
 & $\eta_{\mathrm{b}}$ & 0.83 & 0.84 & $\approx0\%$ \\
 & $\eta_{\mathrm{b}}/\eta_{\mathrm{e}}$ & 0.90 & 0.90 & \\ \hline
 \multirow{3}{*}{\makecell{\SI{22}{\hour}}} & $\eta_{\mathrm{e}}$ & 0.74 & 0.70 & $\approx5\%$ \\
 & $\eta_{\mathrm{b}}$ & 0.55 & 0.45 & $\approx18\%$ \\
 & $\eta_{\mathrm{b}}/\eta_{\mathrm{e}}$ & 0.74 & 0.64 & \\ \hline
 
\end{tabular}
\end{center}
\end{table}

\begin{figure}
    \centering
    \includegraphics[width=9cm]{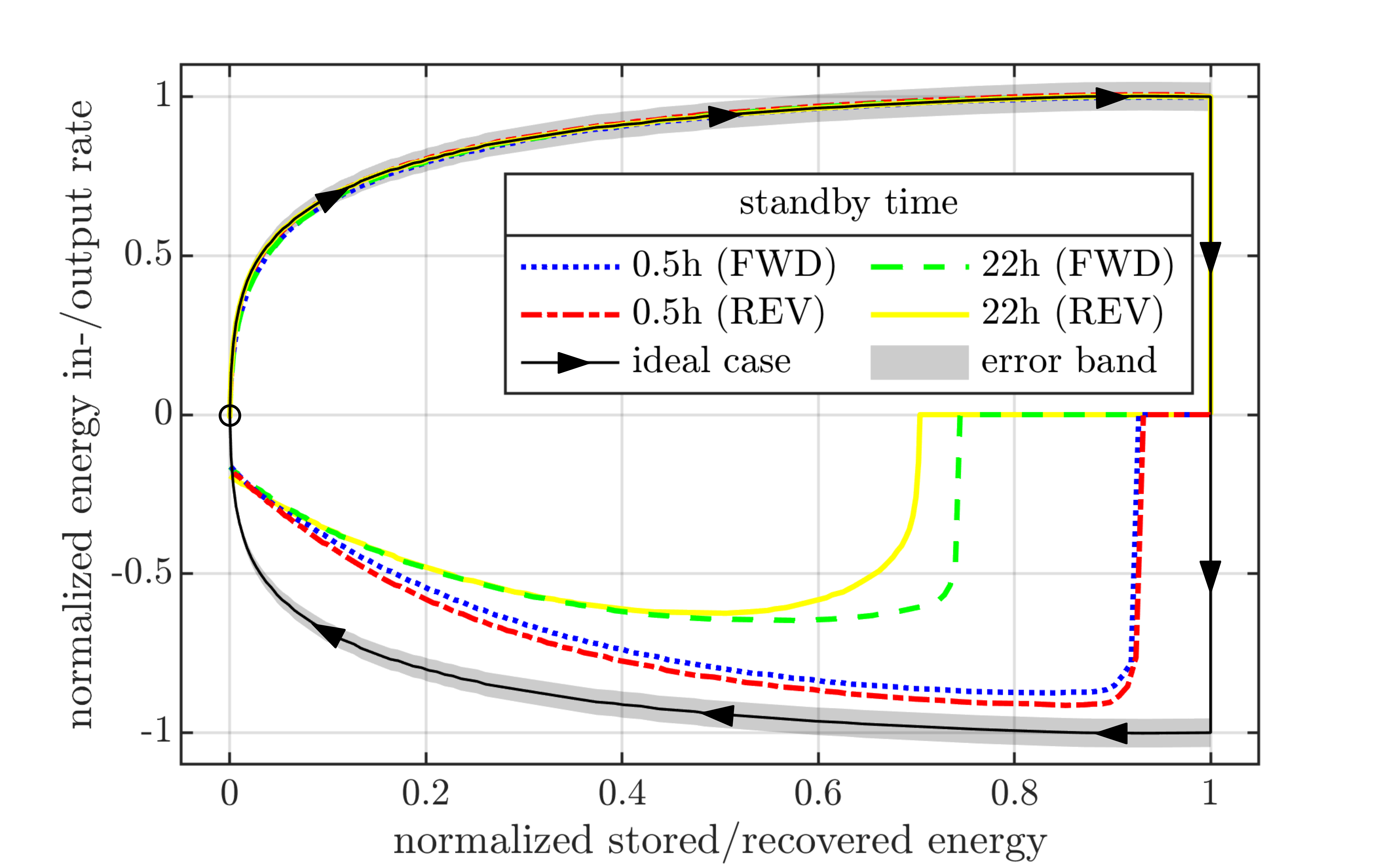}
    \caption{Energy in-/output rate over stored/recovered energy}
    \label{fig:QoverU}
\end{figure}

Now that it has been shown that the influence of natural convection on the standby efficiency of a PBTES depends on the HTF flow direction in the preceding charging period, some important metrics that can be derived from the data are presented and discussed in the following paragraphs. In Figure \ref{fig:QoverU} the normalized energy in- and output rate is plotted over the normalized stored/recovered energy for different experiments. The energy input rate and the stored energy at the end of charging are used for the normalization of the calculated quantities. All experiments start with an empty storage (stored energy equals zero) and an energy input rate of zero. This starting point is marked with a black circle in Figure \ref{fig:QoverU}. During the following charging process the test rig shows similar behaviours for every experiment (black solid line with arrow markers). After two hours of charging the HTF flow is stopped and the storage is switched to standby mode. The recovery of the stored heat (discharging) is started after \SI{0.5}{\hour} in standby for one half of the experiments and after \SI{22}{\hour} for the other half. Considering an ideal storage with zero losses to the surrounding and without thermocline degradation the discharging curve would further follow the solid black line independent on the standby time. In reality the discharging behaviour of the storage is different. The real discharging curves as observed in the experiments are plotted as colored lines in Figure \ref{fig:QoverU}. It can be seen, that the longer the standby period between charging and discharging the storage, the less energy can be recovered. In the case that the storage is switched to discharging mode after just \SI{0.5}{\hour} in standby, which is the minimum standby time that can be realized due to the thermal inertia of the ASU, the energy efficiency is very high ($92\%$, which is similar to observations by Bruch et al. \cite{bruch_experimental_2017}) and even $87\%$ of the theoretically possible maximum energy output rate is reached. These results are similar for both the FWD-case and the REV-case experiments. In the experiments with much longer time periods in standby mode the results for both cases begin to diverge. When the discharging process is started after \SI{22}{\hour} in standby mode, the energy efficiency for the FWD-case is observed to be $74\%$ in contrast to $70\%$ for the REV-case. Similar discrepancies apply to the energy output rate. $65\%$ and $62\%$ of the theoretically possible maximum energy output rate are reached for the FWD- and REV-case respectively.\\

\begin{figure}
    \centering
    \includegraphics[width=9cm]{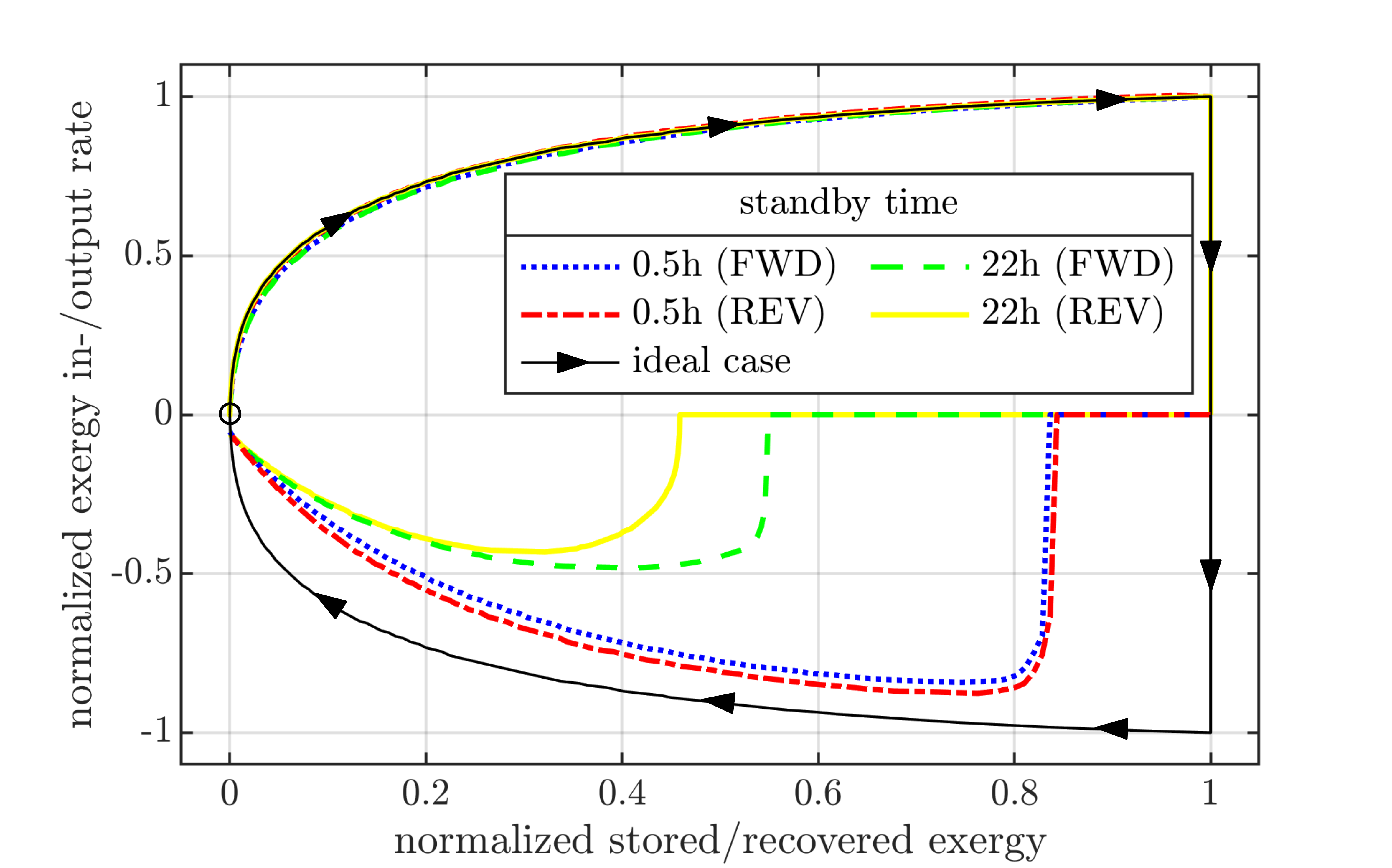}
    \caption{Exergy in-/output rate over stored/recovered exergy}
    \label{fig:BoverU}
\end{figure}

The exergy efficiencies are generally lower than the energy efficiencies due to entropy generation inside the packed bed and finite heat transfer between the storage material and the HTF, which can be seen in Figure \ref{fig:BoverU}. For a standby time of \SI{0.5}{\hour} the exergy efficiency is already as low as $83\%$. Which again is similar for both the FWD-case and the REV-case experiments. In experiments with the longer time periods in standby mode exergy efficiencies of $55\%$ and $45\%$ can be observed for the FWD- and the REV-case respectively. These results are summarized in Table \ref{tab:ratio}. It can be seen that for a standby time of \SI{0.5}{\hour} the efficiencies for both the FWD and the REV case are nearly the same. This means, that both cases are equivalent and no additional losses have to be expected when the flow direction of the HTF in the preceding charging process was from bottom to top. In contrast, losses in the energy efficiency of $5\%$ and losses in exergy efficiency of $18\%$ are observed for a standby time of \SI{22}{\hour} and the same operational parameters. Finally, it should be mentioned, that the calculated efficiencies apply to a PBTES system with similar geometry, insulation and storage temperatures as described in Section \ref{sec:methods} and Table \ref{tab:param}. The results and conclusions drawn in this work are representative for all types of TES that utilize a packed bed as storage material and a gas of a fluid as HTF. However, for a PBTES where one of these parameters is significantly different, the system's efficiencies may differ from the ones presented in this work.

\section{Conclusion}
\label{sec:conclusion}
This study examines the standby efficiency of a packed bed thermal energy storage for different flow directions of the heat transfer fluid in the preceding charging process. The test rig used for the experimental investigations is a lab-scale packed bed thermal energy storage with vertical flow of the heat transfer fluid. The storage tank is filled with LD-slag in the form of irregular shaped, porous rocks with a particle size of $\SIrange{16}{32}{\milli\meter}$ as storage material and it uses air as heat transfer fluid.\\

The results reveal that if the heat stored in a packed bed thermal energy storage is recovered immediately after the charging process, the fluid flow direction in the preceding charging process does not have an impact on the system's energy and exergy efficiencies. However, the efficiencies and thermocline degradation during a long standby period of the storage significantly depend on the flow direction of the heat transfer fluid in the preceding charging process. When the flow direction in the preceding charging process was from bottom to top, entropy generation in the packed bed, and thus exergy losses, due to irreversible effects during a standby period are twice as high compared to the reverse flow direction. This difference can be attributed to natural convection. It could be shown that it takes seven hours for the analyzed system until a fully developed air circulation inside the packed bed is established. For a storage system of which the maximum standby time will be well below these seven hours, the fluid flow direction in the preceding charging process does not have an impact in the storage's standby efficiency. However, if a system is designed for much longer standby periods the charging flow direction should always be from top to bottom. The consequence of charging the storage from bottom instead of from top is a significant decrease in energy and exergy efficiency of $5\%$ and $18\%$ for a 22 hour standby period.\\

The results presented in this work provide detailed information on the efficiency of a packed bed thermal energy storage that can be used as a basis for economic assessments and payback period calculations. Therefore this study contributes to the development of thermal energy storage systems so that they can be deployed to make use of the industry sectors' enormous waste heat potential.

\section*{Acknowledgement}
The authors acknowledge funding support of this work through the research project \textit{5DIndustrialTwin} as part of the Austrian Climate and Energy Fund's initiative Energieforschung (e!MISSION) 6\textsuperscript{th} call (KLIEN/FFG project number 881140). Furthermore, the authors acknowledge TU Wien Bibliothek for financial support through its Open Access Funding programme.

\section*{Declaration of Competing Interest}
The authors declare that they have no known competing financial interests or personal relationships that could have appeared to influence the work reported in this paper.

\appendix


 \bibliographystyle{elsarticle-num} 
 \bibliography{main.bib}

\begin{thebibliography}{10}
\expandafter\ifx\csname url\endcsname\relax
  \def\url#1{\texttt{#1}}\fi
\expandafter\ifx\csname urlprefix\endcsname\relax\def\urlprefix{URL }\fi
\expandafter\ifx\csname href\endcsname\relax
  \def\href#1#2{#2} \def\path#1{#1}\fi

\bibitem{jouhara_editorial_2018}
H.~Jouhara, A.~G. Olabi, Editorial: {Industrial} waste heat recovery, Energy
  160 (2018) 1--2.
\newblock \href {https://doi.org/10.1016/j.energy.2018.07.013}
  {\path{doi:10.1016/j.energy.2018.07.013}}.

\bibitem{papapetrou_industrial_2018}
M.~Papapetrou, G.~Kosmadakis, A.~Cipollina, U.~La~Commare, G.~Micale,
  Industrial waste heat: {Estimation} of the technically available resource in
  the {EU} per industrial sector, temperature level and country, Applied
  Thermal Engineering 138 (2018) 207--216.
\newblock \href {https://doi.org/10.1016/j.applthermaleng.2018.04.043}
  {\path{doi:10.1016/j.applthermaleng.2018.04.043}}.

\bibitem{panayiotou_preliminary_2017}
G.~P. Panayiotou, G.~Bianchi, G.~Georgiou, L.~Aresti, M.~Argyrou,
  R.~Agathokleous, K.~M. Tsamos, S.~A. Tassou, G.~Florides, S.~Kalogirou,
  P.~Christodoulides, Preliminary assessment of waste heat potential in major
  {European} industries, Energy Procedia 123 (2017) 335--345.
\newblock \href {https://doi.org/10.1016/j.egypro.2017.07.263}
  {\path{doi:10.1016/j.egypro.2017.07.263}}.

\bibitem{forman_estimating_2016}
C.~Forman, I.~K. Muritala, R.~Pardemann, B.~Meyer, Estimating the global waste
  heat potential, Renewable and Sustainable Energy Reviews 57 (2016)
  1568--1579.
\newblock \href {https://doi.org/10.1016/j.rser.2015.12.192}
  {\path{doi:10.1016/j.rser.2015.12.192}}.

\bibitem{manente_structured_2022}
G.~Manente, Y.~Ding, A.~Sciacovelli, A structured procedure for the selection
  of thermal energy storage options for utilization and conversion of
  industrial waste heat, Journal of Energy Storage 51 (2022) 104411.
\newblock \href {https://doi.org/10.1016/j.est.2022.104411}
  {\path{doi:10.1016/j.est.2022.104411}}.

\bibitem{gautam_review_2020}
A.~Gautam, R.~P. Saini, A review on sensible heat based packed bed solar
  thermal energy storage system for low temperature applications, Solar Energy
  207 (2020) 937--956.
\newblock \href {https://doi.org/10.1016/j.solener.2020.07.027}
  {\path{doi:10.1016/j.solener.2020.07.027}}.

\bibitem{gautam_review_2020-1}
A.~Gautam, R.~P. Saini, A review on technical, applications and economic aspect
  of packed bed solar thermal energy storage system, Journal of Energy Storage
  27 (2020) 101046.
\newblock \href {https://doi.org/10.1016/j.est.2019.101046}
  {\path{doi:10.1016/j.est.2019.101046}}.

\bibitem{xie_thermocline_2022}
B.~Xie, N.~Baudin, J.~Soto, Y.~Fan, L.~Luo, Thermocline packed bed thermal
  energy storage system: a review, in: M.~Jeguirim (Ed.), Renewable {Energy}
  {Production} and {Distribution}, Vol.~1 of Advances in {Renewable} {Energy}
  {Technologies}, Academic Press, 2022, pp. 325--385.
\newblock \href {https://doi.org/10.1016/B978-0-323-91892-3.24001-6}
  {\path{doi:10.1016/B978-0-323-91892-3.24001-6}}.

\bibitem{esence_review_2017}
T.~Esence, A.~Bruch, S.~Molina, B.~Stutz, J.-F. Fourmigué, A review on
  experience feedback and numerical modeling of packed-bed thermal energy
  storage systems, Solar Energy 153 (2017) 628--654.
\newblock \href {https://doi.org/10.1016/j.solener.2017.03.032}
  {\path{doi:10.1016/j.solener.2017.03.032}}.

\bibitem{stekli_technical_2013}
J.~Stekli, L.~Irwin, R.~Pitchumani, Technical {Challenges} and {Opportunities}
  for {Concentrating} {Solar} {Power} {With} {Thermal} {Energy} {Storage},
  Journal of Thermal Science and Engineering Applications 5~(2) (2013).
\newblock \href {https://doi.org/10.1115/1.4024143}
  {\path{doi:10.1115/1.4024143}}.

\bibitem{bruch_experimental_2014}
A.~Bruch, J.~F. Fourmigué, R.~Couturier, Experimental and numerical
  investigation of a pilot-scale thermal oil packed bed thermal storage system
  for {CSP} power plant, Solar Energy 105 (2014) 116--125.
\newblock \href {https://doi.org/10.1016/j.solener.2014.03.019}
  {\path{doi:10.1016/j.solener.2014.03.019}}.

\bibitem{bruch_experimental_2017}
A.~Bruch, S.~Molina, T.~Esence, J.~F. Fourmigué, R.~Couturier, Experimental
  investigation of cycling behaviour of pilot-scale thermal oil packed-bed
  thermal storage system, Renewable Energy 103 (2017) 277--285.
\newblock \href {https://doi.org/10.1016/j.renene.2016.11.029}
  {\path{doi:10.1016/j.renene.2016.11.029}}.

\bibitem{okello_experimental_2014}
D.~Okello, O.~J. Nydal, E.~J.~K. Banda, Experimental investigation of thermal
  de-stratification in rock bed {TES} systems for high temperature
  applications, Energy Conversion and Management 86 (2014) 125--131.
\newblock \href {https://doi.org/10.1016/j.enconman.2014.05.005}
  {\path{doi:10.1016/j.enconman.2014.05.005}}.

\bibitem{mertens_dynamic_2014}
N.~Mertens, F.~Alobaid, L.~Frigge, B.~Epple, Dynamic simulation of integrated
  rock-bed thermocline storage for concentrated solar power, Solar Energy 110
  (2014) 830--842.
\newblock \href {https://doi.org/10.1016/j.solener.2014.10.021}
  {\path{doi:10.1016/j.solener.2014.10.021}}.

\bibitem{xu_parametric_2012}
C.~Xu, Z.~Wang, Y.~He, X.~Li, F.~Bai, Parametric study and standby behavior of
  a packed-bed molten salt thermocline thermal storage system, Renewable Energy
  48 (2012) 1--9.
\newblock \href {https://doi.org/10.1016/j.renene.2012.04.017}
  {\path{doi:10.1016/j.renene.2012.04.017}}.

\bibitem{rodrigues_stratification_2021}
F.~A. Rodrigues, M.~J. de~Lemos, Stratification and energy losses in a standby
  cycle of a thermal energy storage system, International Journal of Energy for
  a Clean Environment 22 (2021) 1--32.

\bibitem{yang_study_2019}
B.~Yang, F.~Bai, Y.~Wang, Z.~Wang, Study on standby process of an air-based
  solid packed bed for flexible high-temperature heat storage: {Experimental}
  results and modelling, Applied Energy 238 (2019) 135--146.
\newblock \href {https://doi.org/10.1016/j.apenergy.2019.01.073}
  {\path{doi:10.1016/j.apenergy.2019.01.073}}.

\bibitem{Koller_milp_2019}
M.~Koller, R.~Hofmann, H.~Walter, {MILP} model for a packed bed sensible
  thermal energy storage, Computers \& Chemical Engineering 125 (2019) 40--53.
\newblock \href {https://doi.org/10.1016/j.compchemeng.2019.03.007}
  {\path{doi:10.1016/j.compchemeng.2019.03.007}}.

\bibitem{gasia_review_2017}
J.~Gasia, L.~Miró, L.~F. Cabeza, Review on system and materials requirements
  for high temperature thermal energy storage. {Part} 1: {General}
  requirements, Renewable and Sustainable Energy Reviews 75 (2017) 1320--1338.
\newblock \href {https://doi.org/10.1016/j.rser.2016.11.119}
  {\path{doi:10.1016/j.rser.2016.11.119}}.

\bibitem{cabeza_1_2021}
L.~F. Cabeza, I.~Martorell, L.~Miró, A.~I. Fernández, C.~Barreneche, L.~F.
  Cabeza, A.~I. Fernández, C.~Barreneche, 1 - {Introduction} to thermal energy
  storage systems, in: L.~F. Cabeza (Ed.), Advances in {Thermal} {Energy}
  {Storage} {Systems} ({Second} {Edition}), Woodhead {Publishing} {Series} in
  {Energy}, Woodhead Publishing, 2021, pp. 1--33.
\newblock \href {https://doi.org/10.1016/B978-0-12-819885-8.00001-2}
  {\path{doi:10.1016/B978-0-12-819885-8.00001-2}}.

\bibitem{schwarzmayr_development_2022}
P.~Schwarzmayr, F.~Birkelbach, L.~Kasper, R.~Hofmann, Development of a
  {Digital} {Twin} {Platform} for {Industrial} {Energy} {Systems}, in:
  Accelerated {Energy} {Innovations} and {Emerging} {Technologies}, Vol.~25,
  Energy Proceedings, Cambridge, USA, 2022.
\newblock \href
  {https://doi.org/https://doi.org/10.46855/energy-proceedings-9974}
  {\path{doi:https://doi.org/10.46855/energy-proceedings-9974}}.

\bibitem{merker_konvektive_1987}
G.~P. Merker, Konvektive {Wärmeübertragung}, Wärme- und {Stoffübertragung},
  Springer Berlin Heidelberg, Berlin, Heidelberg, 1987.
\newblock \href {https://doi.org/10.1007/978-3-642-82890-4}
  {\path{doi:10.1007/978-3-642-82890-4}}.

\end{thebibliography}


\printnomenclature
\nomAcro[tes]{TES}{thermal energy storage}
\nomAcro[pbtes]{PBTES}{packed bed thermal energy storage}
\nomAcro[htf]{HTF}{heat transfer fluid}
\nomAcro[ld]{LD}{Linz-Donawitz}
\nomAcro[asu]{ASU}{air supply unit}
\nomAcro[fwd]{FWD}{forward}
\nomAcro[rev]{REV}{reverse}
\nomAcro[eu]{EU}{European Union}
\nomAcro[rtd]{RTD}{Resistance temperature detector}

\nomRoman[t]{$t$}{time in \si{\second}}
\nomRoman[T]{$T$}{temperature in \si{\kelvin}}
\nomRoman[s]{$S$}{entropy in \si{\joule\per\kelvin}}
\nomRoman[sdot]{$\dot{S}$}{entropy flow rate in \si{\watt\per\kelvin}}
\nomRoman[q]{$Q$}{heat in \si{\joule}}
\nomRoman[qdot]{$\dot{Q}$}{heat flow rate from/to HTF or surrounding in \si{\watt}}
\nomRoman[c]{$c$}{specific heat of storage material \si{\joule\per\kilo\gram\per\kelvin}}
\nomRoman[cp]{$c_p$}{specific heat at constant pressure for heat transfer fluid in \si{\joule\per\kilo\gram\per\kelvin}}
\nomRoman[m]{$m$}{mass of storage material in \si{\kilo\gram}}
\nomRoman[mdot]{$\dot{m}$}{mass flow rate of HTF in \si{\kilo\gram\per\second}}
\nomRoman[k]{$k$}{heat transfer coefficient in \si{\watt\per\square\meter\per\kelvin}}
\nomRoman[a]{$A$}{surface area in \si{\square\meter}}
\nomRoman[u]{$U$}{internal energy in \si{\joule}}
\nomRoman[b]{$B$}{exergy in \si{\joule}}
\nomRoman[bdot]{$\dot{B}$}{exergy flow rate in \si{\watt}}
\nomRoman[h]{$h$}{specific enthalpy in \si{\joule\per\kilo\gram}}
\nomRoman[n]{$n$}{number of vertical volume sections $n=9$}

\nomGreek[n]{$\eta$}{efficiency}
\nomGreek[l]{$\lambda$}{thermal conductivity in \si{\watt\per\meter\per\kelvin}}

\nomSub[i]{$i$}{volume element index}
\nomSub[sys]{$sys$}{system}
\nomSub[tb]{$t/b$}{top/bottom}
\nomSub[gen]{$gen$}{generation}
\nomSub[s]{$s$}{surface}
\nomSub[lat]{$lat$}{lateral}
\nomSub[amb]{$amb$}{ambient}
\nomSub[htf]{$HTF$}{heat transfer fluid}
\nomSub[in]{$in$}{inlet}
\nomSub[out]{$out$}{outlet}
\nomSub[ref]{$ref$}{reference ($T_{\mathrm{ref}} = \SI{295.15}{\kelvin}$)}
\nomSub[e]{$e$}{energy}
\nomSub[b]{$b$}{exergy}
\nomSub[recover]{$recover$}{recover}
\nomSub[charge]{$charge$}{charge}
\nomSub[q]{$Q$}{heat}
\nomSub[rel]{$rel$}{relative}





\end{document}